# Robust B-exciton emission at room temperature in few-layers of MoS$_2$:Ag nanoheterojunctions embedded into a glass matrix

Abdus Salam Sarkar[1*], Ioannis Konidakis[1], Ioanna Demeridou[1,2], Efthymis Serpetzoglou[1,2], George Kioseoglou[1,3], and Emmanuel Stratakis[1,2*]

[1]Institute of Electronic Structure and Laser, Foundation for Research and Technology-Hellas, Heraklion, 700 13 Crete, Greece.

[2]Physics Department, University of Crete, Heraklion, 710 03 Crete, Greece.

[3]Department of Materials Science and Technology, University of Crete, Heraklion, 710 03 Crete, Greece.

**\*Email**: salam@iesl.forth.gr; stratak@iesl.forth.gr

**Abstract**

Tailoring the photoluminescence (PL) properties in two-dimensional (2D) molybdenum disulfide ($MoS_2$) crystals using external factors is critical for its use in valleytronic, nanophotonic and optoelectronic applications. Although significant effort has been devoted towards enhancing or manipulating the excitonic emission in $MoS_2$ monolayers, the excitonic emission in few-layers $MoS_2$ has been largely unexplored. Here, we put forward a novel nano-heterojunction system, prepared with a non-lithographic process, to enhance and control such emission. It is based on the incorporation of few-layers $MoS_2$ into a plasmonic silver metaphosphate glass ($AgPO_3$) matrix. It is shown that, apart from the enhancement of the emission of both A and B excitons, the B-excitonic emission dominates the PL intensity. In particular, we observe an almost six-fold enhancement of the B exciton emission, compared to control $MoS_2$ samples. This enhanced PL at room temperature is attributed to an enhanced exciton-plasmon coupling and it is supported by ultrafast time-resolved spectroscopy that reveals plasmon-enhanced electron transfer that takes place in Ag nanoparticles-$MoS_2$ nanoheterojunctions. Our results provide a great avenue to tailor the emission properties of few-layers $MoS_2$, which could find application in emerging valleytronic devices working with B excitons.

**Introduction**

Two-dimensional (2D) Transition Metal Dichalcogenides (TMDs) provide an appealing platform for emerging atomic scale research in nanophotonic and optoelectronic applications.[1-4] Monolayer molybdenum disulfide ($MoS_2$), in particular, gains considerable attention due to its direct band gap and potential integration with other nanostructures to form nanoscale van der Waals heterojunctions with intriguing physical and optical properties.[5] Indeed, it has been shown that the optical properties of $MoS_2$ monolayers, such as photoluminescence (PL), can be manipulated through its coupling with nanomaterials of various dimensionalities. In particular, zero dimensional (0D) quantum dots and nanoparticles,[6, 7] one-dimensional (1D) nanowires and nanorods,[8-10] as well as other 2D materials[5, 11] had been combined with monolayer $MoS_2$ to manipulate its emission intensity and/or quantum yield. Besides this, polymeric spacing,[12] defect engineering,[13] doping,[14] and chemical modification[15] approaches were employed to manipulate the emission properties. However, monolayer $MoS_2$ suffers from low intrinsic photoluminescence (PL) quantum yield (0.01 to 0.6%), dominated by the A- excitonic emission, due to its sub nanometer thickness and defect density mediated nonradiated recombination.[1] The low PL yield was overcome (more than 95%) with chemical treatment by an organic superacid.[16] In contrast to a monolayer $MoS_2$, few layers of $MoS_2$ have several orders of magnitude lower PL quantum yield.[1] On the other hand, few-layers $MoS_2$, as an indirect semiconductor, have significantly larger optical density, which enhances its external quantum efficiency.[17] Owing to this advantage, research on the PL properties in few layers $MoS_2$ has received significant attention. For example, metallic and other nanostructures[5, 7] were used to manipulate the A-excitonic emission in few layers of $MoS_2$.[18] However, this approach has only been limited to the enhancement of the A-excitonic emission.

On the other hand, transparent thermoplastic glasses (TTG) were extensively used for homogeneous incorporation of 2D layered materials. However, the relevant studies were limited to measure the nonlinear optical response of the embedded 2D nanoflakes.[19, 20] On a rather different manner photonic crystal cavities,[21-23] as well as Mie-resonant metasurfaces,[6] have been employed to tailor the optical properties of $MoS_2$. Similar to the case of nanostructures, the manipulation of PL emission has been only limited to A-exciton. Mikkelsen and co-workers[24, 25] were the first who carried out a systematic study to manipulate the B-excitonic emission of a single-layer of $MoS_2$. However, the study of the emission properties was limited to the ground A-exciton state. Nevertheless, a detailed investigation of the B-exciton state in the ultrafast regime is crucial to shed light on the physical phenomena that take place.

In this study, we present the development of a nanohybrid heterojunction system composed of few layers of $MoS_2$ embedded into a silver metaphosphate glass ($AgPO_3$), as a means to enhance and control the $MoS_2$ exciton emission. The selection of $AgPO_3$ glass as a host matrix is prompted by several reasons: First, its transparency in most of the visible range (Fig. S1) enables the full exploitation of the $AgPO_3$:$MoS_2$ photoluminescence properties towards various nanophotonic applications.[26] Moreover, the presence of silver nanoparticles (NPs) within the glass matrix gives rise to interesting optical phenomena that can be exploited towards enhancing and manipulating the PL properties of the incorporated $MoS_2$ layers. Finally, the $AgPO_3$ glass exhibits a very low glass transition temperature of 192 °C, which is indicative of its soft nature. As a consequence, the $MoS_2$ integration process is performed at low temperatures, suitable to avoid any oxidation. On top of that, an advanced 2D exciton-plasmon system composed of a few-layer TMD integrated with a semiconducting metal-phosphate glass is realized. It is shown that the layered TMDs create nanoscale van der Waals heterojunctions with the metallic nanostructures of the glass, which can be exploited to tailor light-matter interactions at the nanoscale.

**Results**

*Fabrication and characterization of MoS$_2$ and nanoheterojunctions.* The MoS$_2$ flakes were obtained by liquid exfoliation (see methods).[27] The lateral dimensions of the MoS$_2$ nanoflakes, as determined by SEM imaging, were found to lie within the micrometer range (Fig. S2a), while, the average thickness measured by AFM was ~4 nm (Fig. S2 b, c). A schematic representation of the composite glass, comprising numerous AgPO$_3$:MoS$_2$ nano-heterojunctions is illustrated in Scheme 1 (see methods).

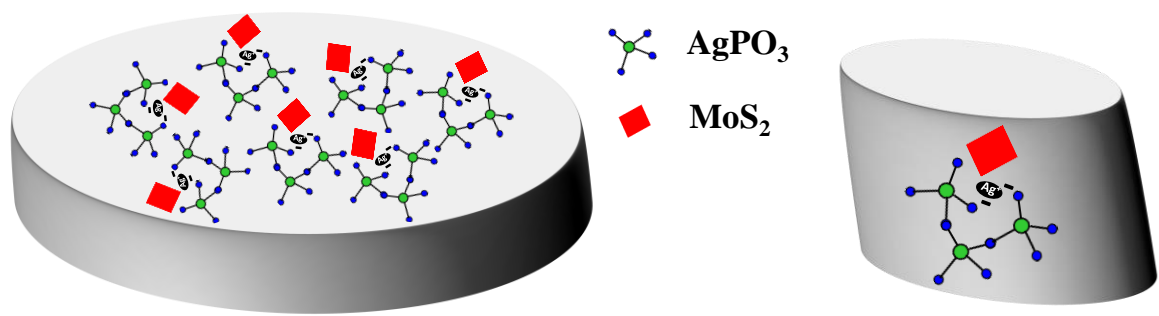

**Scheme 1.** Schematic representation of the composite transparent silver metaphosphate glass (AgPO$_3$) matrix incorporating the MoS$_2$ flakes. The right panel depicts a single AgPO$_3$:MoS$_2$ nanoheterojunction.

*Optical spectroscopy.* Absorption spectroscopy was employed to confirm the formation of nano-heterojunctions between MoS$_2$ and AgPO$_3$. The pristine AgPO$_3$ glass exhibits two characteristic peaks at 2.0 and 2.5 eV (Fig. S3, a and b), which correspond to the Ag plasmonic bands and are attributed to a bimodal distribution of isolated nanoparticles or their clusters attained due to the phosphate matrix. Another reason of the emergence of the plasmonic band at 2.0 eV is the clustering/agglomeration of Ag NPs observed. Indeed, as the effective nanoparticles size increases, a corresponding red shift in the plasmon band occurs. This is also indicated by the broad plasmonic band centered at 2.0 eV. The absorption spectrum of bare $MoS_2$ flakes exhibits the two characteristic excitonic peaks at 1.84 eV (A-exciton) and 2.03 eV (B-exciton) respectively (Fig. 1a, red line).[1] Both peaks were also present in the absorption spectrum of the $AgPO_3$:$MoS_2$ heterojunctions of composite matrix (green line of Fig. 1a), i.e. in which the $MoS_2$ is incorporated within the glass. At the same time, the absorption intensity of $AgPO_3$:$MoS_2$, is enhanced compared to the pristine $AgPO_3$. The small blue shift of the A- and B-exciton peak positions by 13meV and 17meV, respectively, is due to the change of the dielectric environment rather than any oxidation process. The integration of $MoS_2$ within $AgPO_3$ glass took place at 170°C, which is below the glass transition ($T_G$), and thus hard to cause significant oxidation of the phosphate network. Instead, the modification of the dielectric constant, from that of the solvent to the higher dielectric constant of the surrounding glass matrix, could be the reason for the observed shift of the exciton states. It is noted that these findings are not observed when the $MoS_2$ flake is positioned on the surface of $AgPO_3$ glass, i.e. $AgPO_3$/$MoS_2$ spectrum in Fig. 1a (blue line)

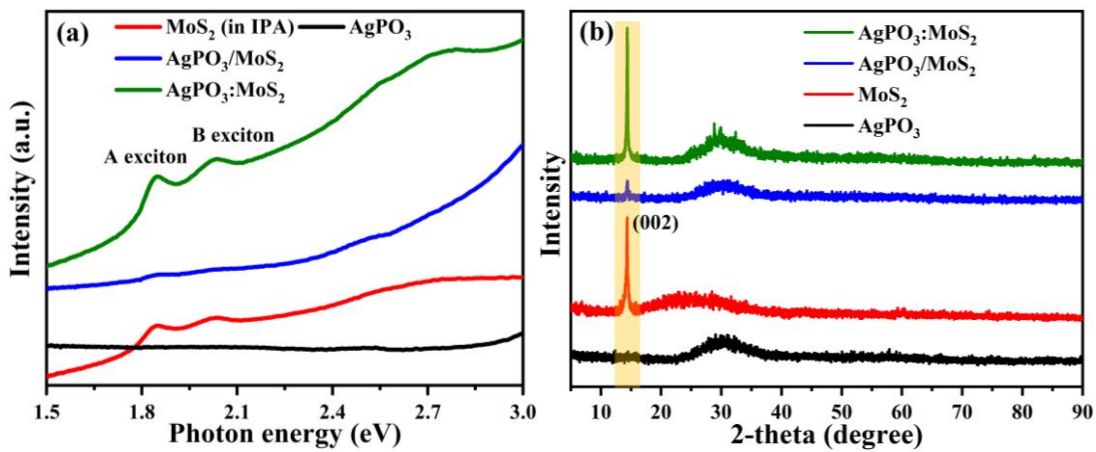

**Figure 1. Spectroscopic characterizations of MoS2 and their heterojunctions (a)** Optical absorption spectra and **(b)** X-ray diffraction pattern, of $MoS_2$ ($MoS_2$ on glass substrate), $AgPO_3$, $AgPO_3$/$MoS_2$ ($MoS_2$ on $AgPO_3$), and $AgPO_3$:$MoS_2$ ($MoS_2$ embedded into $AgPO_3$).

It is widely acknowledged that the trigonal prismatic phase (2H-phase) integrity plays a crucial role in PL emission of exfoliated $MoS_2$. Aiming to identify the phase integrity in $MoS_2$ flakes dispersed into $AgPO_3$, a series of structural studies have been carried out. In particular, the X-ray diffraction pattern of $MoS_2$ (Fig. 1b) exhibits a strong peak located at 14.34º, corresponding to the (002) plane, which agrees well with the hexagonal $MoS_2$.[28, 29] Besides this, the examination of the $AgPO_3/MoS_2$ and $AgPO_3$:$MoS_2$ matrix showed a primary peak at 2θ~14.38º, indicating that the liquid exfoliation of few layers did not change the $MoS_2$ structure. Since the peak position in XRD is not changing for both structures (Fig. S3), there is no significant strain induced when $MoS_2$ is incorporated into the phosphate matrix. However, a reduction in the full width at half maximum (FWHM) has been observed (Fig. S3c and Table S1) and this could be due to the variation in the microstructure, the grain distortion, and dislocation density of the crystal.[30, 31] The possibility that higher crystallinity may have reduce the FWHM of $MoS_2$ (because of the heating process) was excluded by performing a controlled experiment in $MoS_2$ treated under the same conditions used for the fabrication of $AgPO_3$:$MoS_2$. No changes in the peak position as well as in the FWHM were observed (Fig. S3b).

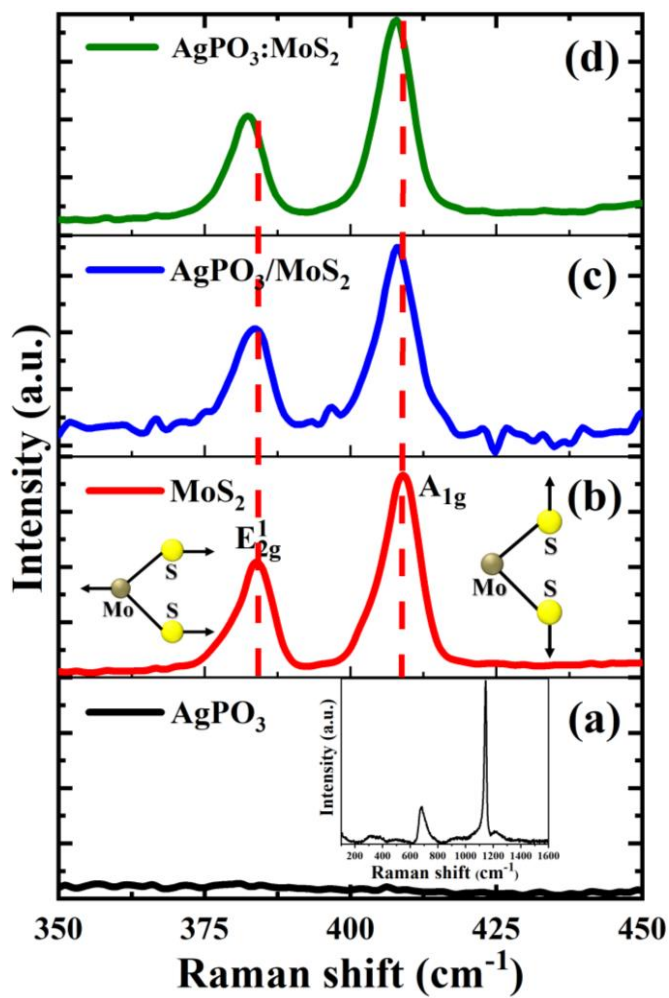

**Figure 2.** Raman spectra of **(a)** AgPO$_3$ glass, **(b)** MoS$_2$ flakes on Si (Si/MoS$_2$), **(c)** MoS$_2$ on AgPO$_3$ (AgPO$_3$/MoS$_2$) and **(d)** MoS$_2$ embedded into AgPO$_3$ (AgPO$_3$:MoS$_2$). The inset presents the Raman spectrum of the pristine AgPO$_3$ glass in a wide range.

In addition, the Raman spectra (Fig. 2) of AgPO$_3$:MoS$_2$ composite glass were obtained and compared with that of a bare MoS$_2$. The Raman spectra of bare MoS$_2$ depict two characteristic peaks at 382.83 and 407.12 cm$^{-1}$, corresponding to the in-plane ($E_{2g}^1$) and out of plane (A$_{1g}$) vibrational modes. The Raman frequency difference ($\Delta\omega=\omega(A_{1g}) - \omega(E_{2g}^1)$) between these two modes dependents on the number of layers, which is used to determine the MoS$_2$ thickness[32, 33]; this difference ($\Delta\omega$) is measured to be 24-25 cm$^{-1}$, indicating that the MoS$_2$ flakes have several layers.[32-34] This number is found to be similar for all the samples studied (Fig. S4). Besides this, the full width at half maximum (FWHM) of $E_{2g}^1$ is ~5.67 cm$^{-1}$, suggests good crystallinity of the exfoliated MoS$_2$.[35,36] Furthermore, a small red shift of about 2 cm$^{-1}$ in both Raman modes has been observed in AgPO$_3$:MoS$_2$[36]. This shift is unlikely to be due to strain since it is the same for both modes and not only for the in-plane one (a signature of induced strain in the system). The inset of Fig. 2 also shows the obtained Raman spectrum of the pristine AgPO$_3$ glass, i.e. prior to any MoS$_2$ incorporation. The spectrum of AgPO$_3$ glass exhibits a major band at around 1142 cm$^{-1}$, whereas a broader band at ~ 675 cm$^{-1}$ is also present. The first Raman signature is attributed to the symmetric stretching vibration of terminal PO$_2^-$ groups, $v_s(PO_2^-)$, while the latter features originates from the symmetric stretching movement of P-O-P bridges within the phosphate backbone, $v_s$(P-O-P).[37, 38] μ-photoluminescence (μ-PL) spectroscopy was employed to investigate the emission properties of MoS$_2$ flakes embedded into the AgPO$_3$ matrix. Fig. 3a presents the steady state PL spectra of all the samples. The red, black, and blue curves correspond to the PL spectra of AgPO$_3$, Si/MoS$_2$, and AgPO$_3$:MoS$_2$, respectively. As a reference, we first measured the intrinsic PL spectra of MoS$_2$ flakes deposited on Si substrates, using excitation energy of 2.28 eV (543 nm); considering that the direct excitonic transition is weakened with increasing the layer number in MoS$_2$, a broad emission peak with weak intensity was observed (Fig. S4).[18, 39] It is notable that the MoS$_2$ spectrum has been dramatically changed upon its incorporation in AgPO$_3$. Indeed, the spectrum exhibited two well-defined emission peaks at 1.89 and 2.05 eV, corresponding to A- and B- excitonic transition of MoS$_2$, respectively. At the same time the PL emission is significantly enhanced, corresponding to a 5- and 6- fold enhancement in the A- and B- exciton peak intensities (Fig. 3b). It should be noted that the PL spectrum of AgPO$_3$ glass, presented in Fig. 3a shows no emission within the MoS$_2$ excitonic emission range. Notably, besides the enhancement in the PL intensity, and contrary to the conventional PL properties of few-layered MoS$_2$, a dominant B-excitonic peak is observed in the AgPO$_3$:MoS$_2$ emission spectra (Fig. 3b). The corresponding intensity ratio of B and A excitons ($I_B/I_A$) equals to 1.2.

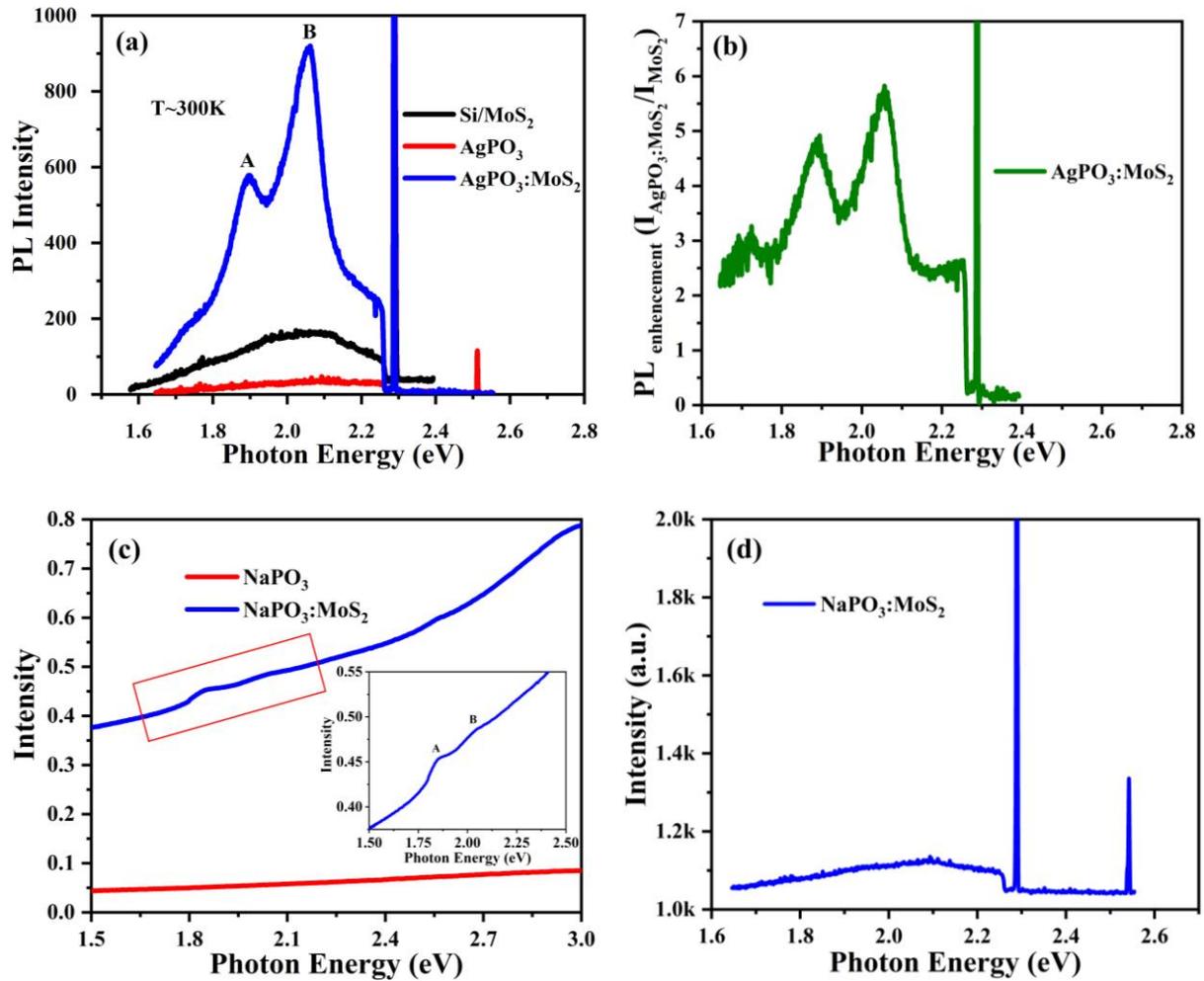

**Figure 3. Room temperature spectroscopic characteristics. (a)** PL emission spectrum of Si/MoS$_2$, AgPO$_3$, and AgPO$_3$:MoS$_2$. **(b)** PL intensity ratio of AgPO$_3$:MoS$_2$ and Si/MoS$_2$ samples. **(c)** absorption spectra of NaPO$_3$, and NaPO$_3$:MoS$_2$ (Inset: Magnified spectra of MoS$_2$ excitonic peaks in NaPO$_3$:MoS$_2$) and **(d)** PL emission spectrum of NaPO$_3$:MoS$_2$.

To understand the effect of AgPO$_3$ matrix on the PL properties, we have investigated the internal structure of AgPO$_3$ by means of TEM microscopy. TEM studies reveal a bimodal size distribution of Ag nanoparticles with dominant average sizes of 8.4 nm and 14.5 nm while a broad size variation was also observed (Fig. S5c). The elemental composition of Ag was confirmed by EDX mapping (Fig. S5c). In this context, the large enhancement of $MoS_2$ PL intensity observed in the $AgPO_3$:$MoS_2$ system can be attributed to the localized surface plasmon effect due to the presence of Ag NPs. In order to provide concrete evidence that the observed enhancement of $AgPO_3$:$MoS_2$ PL intensity is induced by the presence of surface plasmon of Ag particles, we prepare a similar $NaPO_3$:$MoS_2$ heterojunction, i.e. in which the silver is replaced by sodium, while the phosphate glass network remains unchanged.

Fig. 3c shows that the steady state absorption spectra of $NaPO_3$:$MoS_2$ glass exhibits the two characteristic features at 1.84 and 2.03 eV, which are attributed to intrinsic A- and B-excitonic peaks of $MoS_2$, respectively. Moreover, Raman spectroscopy reveals the presence of a few $MoS_2$ layers within the fabricated $NaPO_3$:$MoS_2$ composite glass (Fig. S6). The $NaPO_3$ glass exhibits its own characteristic vibrational modes at around 1155 cm$^{-1}$ and ~ 681 cm$^{-1}$, respectively. Contrary to the case of $AgPO_3$:$MoS_2$, no enhancement of the $MoS_2$ PL is found for the $NaPO_3$:$MoS_2$ system. Indeed, the corresponding room temperature PL spectrum of $NaPO_3$:$MoS_2$ (Fig. 3d), displays only a very broad and extremely weak emission in the range of the direct A- and B-excitonic transitions. It is therefore clear that the remarkable enhancement on the emission properties of the $AgPO_3$:$MoS_2$ is induced by the presence of Ag NPs and their plasmon resonance.

The surface plasmon resonance ($\omega_{LSPR}$), which can be tuned by varying the size and shape of the nanostructures and surrounding dielectric medium, is known to strongly modify the excitonic emission.[25] In particular, the plasmon resonance was tuned by changing the nanostructure size, which enhanced the intrinsically weakly emitting B exciton of a $MoS_2$ flake.[25] We investigated how the silver plasmon resonance affects the emission properties of the developed $MoS_2$ glass heterojunctions upon changing silver content and particle size in the glass. To this aim, an additional glass-MoS$_2$ heterojunction was fabricated upon employing the ternary silver-rich 0.3AgI-0.7AgPO$_3$ glass instead of the binary AgPO$_3$ glass, i.e. for the development of 0.3AgI-0.7AgPO$_3$:MoS$_2$ architecture. In one of our previous studies it was demonstrated that the incorporation of AgI in the AgPO$_3$ glass results to the agglomeration of silver nanoparticles for the formation of larger silver phases,[26] while the phosphate network connectivity remains unaffected. Namely, it was reported that for the aforementioned nominal glass composition silver clusters (larger particles formed from the agglomeration of many nanoparticles) with an average size of 2.78 μm are formed while randomly positioned along the glass network (Fig. S7). The absorption spectrum of bare 0.3AgI-0.7AgPO$_3$ exhibited the broad feature of absorption with a hump at ~2.47 eV (Fig. S7).

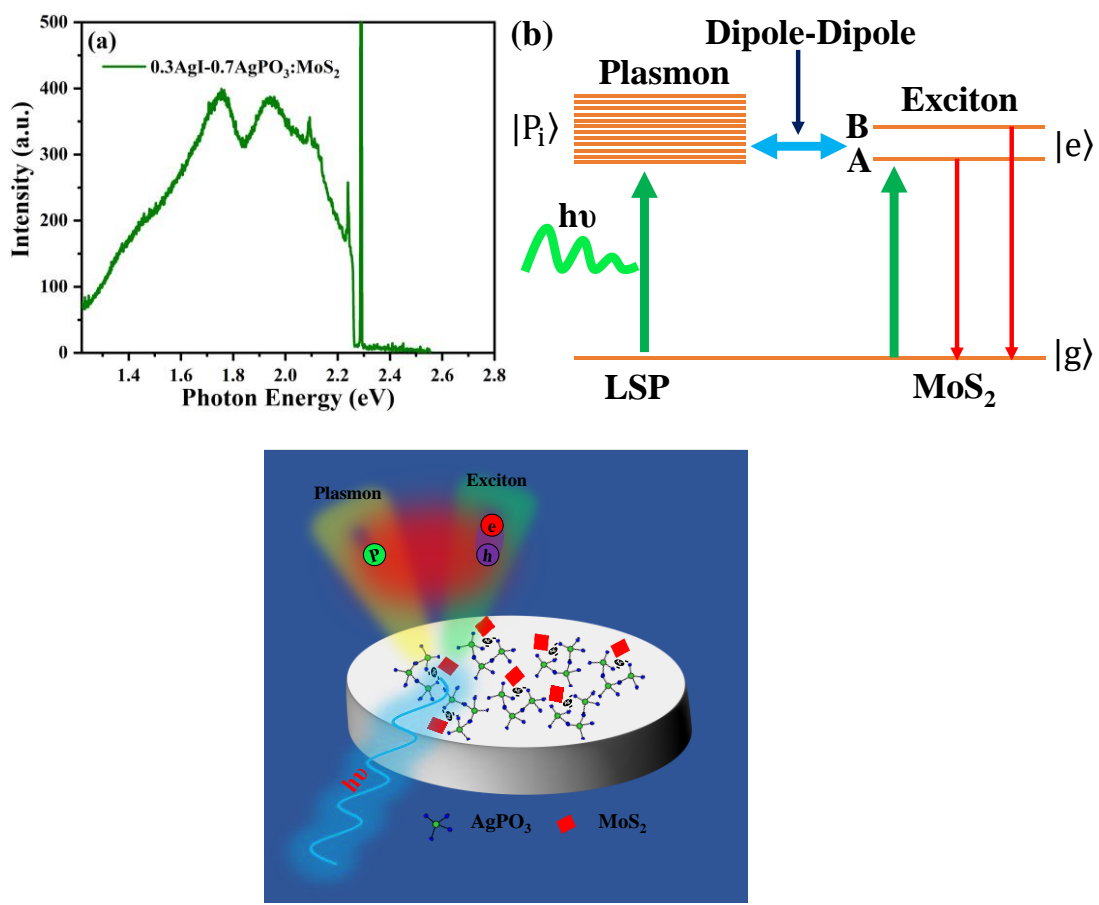

**Figure 4. (a)** Room temperature PL emission spectrum of 0.3AgI-0.7AgPO$_3$:MoS$_2$ and **(b)** Schematic of the mechanism illustrating the exciton-plasmon coupling in MoS$_2$ and AgPO$_3$ via dipole-dipole interaction.

We now consider the effect of these large silver phases on the exciton emission properties of the so-formed 0.3AgI-0.7AgPO$_3$:MoS$_2$ heterojunctions. The measurement conditions were kept identical to these employed for the AgPO$_3$:MoS$_2$ measurements. However, the MoS$_2$ emission spectrum has changed upon its incorporation in 0.3AgI-0.7AgPO$_3$ when compared to AgPO$_3$ (Fig. 4). Specifically, the PL spectrum exhibits two well defined excitonic emission peaks at 1.75 eV and 1.94 eV, corresponding to A- and B- excitonic transitions of MoS$_2$, respectively. The obtained shift in B exciton peak position has been appeared to be 110 meV. These excitonic peaks are red shifted when compared to the considerably smaller size Ag NPs of the AgPO$_3$:MoS$_2$ architecture, and their intensities are almost identical. The corresponding intensity ratio of B- and A- excitons ($I_B/I_A$) is 0.99 in this case, compared to 1.2 of AgPO$_3$:MoS$_2$. Huang et al.[25] have observed similar red-shifts in the peak position for the A and B excitons, due to the nanocavity resonance controlled by the size of plasmonic nanostructures. Thus, the enhancements in the peak intensities and positions of the excitonic transitions are strongly modified by the plasmon nanostructure size in metaphosphate glass.

*Ultrafast carrier dynamics in MoS$_2$ nanoheterojunctions.* Finally, in order to further shed light on the observed enhancement in A- and B- excitonic emissions we investigated the corresponding charge carrier relaxation dynamics by means of ultrafast pump probe time-resolved transient absorption spectroscopy (TAS) (Fig. S8).[40, 41] Fig. 5a presents optical density ($\Delta$OD) vs. wavelength plots at various time delays following photo-excitation of the AgPO$_3$:MoS$_2$ glass using a pump fluence of 2.8 mJ cm$^{-2}$. Fig. 5b shows the photo-bleaching recovery kinetics of A- and B- exciton states at 680 nm (1.82 eV) and 620 nm (2 eV), respectively. In agreement to previous findings,[42] it is observed that the formation of the A exciton is around 0.5 ps slower when compared to that of the B exciton. In particular, the maximum photo-bleaching of the latter is obtained instantly upon photo-excitation at almost 0 ps. This finding is attributed to the electron-hole cooling time from the upper valence state to the lower valence state within the valence band of the $MoS_2$.[42]

Furthermore, upon following typical exponential fittings, we were able to distinguish the physical mechanisms of A- and B- exciton decay dynamics.[42, 43] For the latter exciton state, the bi-exponential fitting procedure based on the equation $y = y_o + A_1 \exp(-x/\tau_1) + A_2 \exp(-x/\tau_2)$, clearly reveals the presence of two distinct times. Namely, an ultrafast component ($\tau_1$) of around 0.5 ps that corresponds to electron transfer from the $MoS_2$ exciton to $AgPO_3$, and a slightly slower time component ($\tau_2$) of around 2 ps that is attributed to carrier-carrier interactions.[42, 43] Rather differently, in the case of A-exciton the ultrafast time component is apparently absent, a finding that implies no electron transfer from the lower energy excitation state towards the metallic particles of the hosting glass. The fast charge transfer present only in the B-exciton, explains why the PL enhancement for the B-exciton is only 6-fold and comparable to the 5-fold observed for the A-exciton (Fig. 3b). There are two effects taking place in the B-exciton during the photoexcitation process i) a PL enhancement due to the efficient dipole coupling of exciton-plasmon and ii) a fast charge transfer from the $MoS_2$ to the AgPO3. These effects are antagonistic and lead to the observed enhancement.

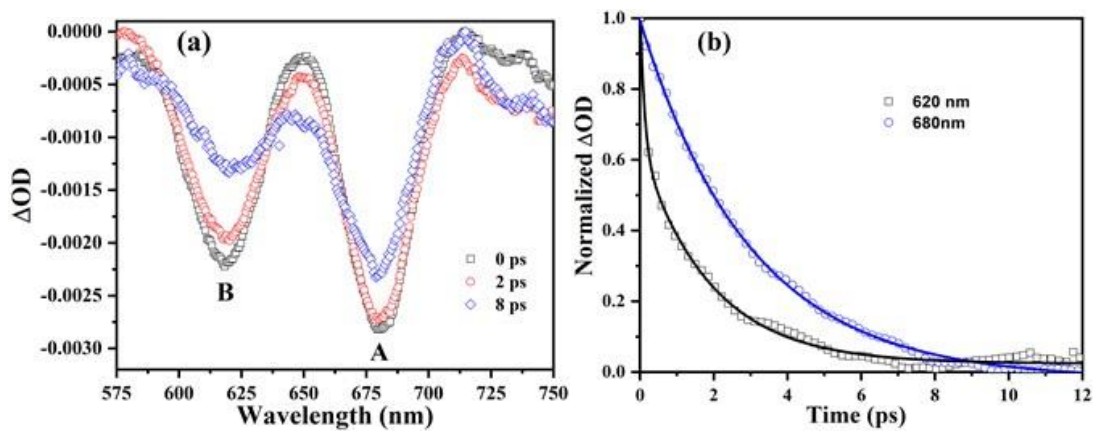

**Figure 5. Transient absorption study of $AgPO_3$:$MoS_2$ nanoheterojunctions. (a)** Optical density (ΔOD) vs. wavelength plots at various time delays following photoexcitation of AgPO$_3$-MoS$_2$ glass at 1026 nm with a pump fluence of 2.8 mJ cm$^{-2}$. **(b)** Normalized transient bleach kinetics (symbols) of the A- and B-excitons at 680 and 620 nm, respectively. The solid lines represent the corresponding decay exponential fits.

Based on the aforementioned results, the plasmon coupling in silver based glasses and MoS$_2$ heterojunction can be facilitated by either electromagnetic field enhancement due to localized surface plasmon (LSP) effect in Ag NPs, and/or via efficient charge injection between the Ag NPs and MoS$_2$ flakes.[18, 44, 45] Screening and scattering effects due to the presence of metallic NPs could also slightly influence the PL intensities.[18] Moreover, heating and strain effects induced by the glass matrix could also contribute to the change in the PL spectrum observed.[39, 46, 47] However, such effects should have negligible influence on the PL enhancement in our case due to the inherent indirect band gap,[18] coupled with large thermal conductivity[48, 49] of the few-layered MoS$_2$ flakes. It can thus be concluded that the exciton (in MoS$_2$)-plasmon (in Ag) coupling (or LSP) is the most plausible explanation for the observed enhancement in PL intensity in Ag based heterojunctions.

To this date, the investigation of surface plasmon induced PL enhancement is only reported in the case of monolayer TMDs.[24, 50] In particular, it is observed that the exciton-plasmon coupling is greatly influenced by the contact area between the plasmon nanostructure and 2D material.[9, 47, 51, 52] In our case, it is obvious that the AgPO$_3$ glass comprises plenty of nanoheterojunctions among MoS$_2$ and Ag NPs, than in the 0.3AgI-0.7AgPO$_3$ glass. The AgPO$_3$ glass contained smaller diameter nanostructures than 0.3AgI-0.7AgPO$_3$ (2.78 μm nanocluster), which can significantly enlarge the contact area and the spatial distribution of the localized electromagnetic field. Besides this, the exact modification of the A- and B- excitonic peaks should strongly depend on the nanoheterojunctions cavity resonance (dipole-dipole interaction), which is controlled by nanostructure size.[24] The steady state photoluminescence enhancement (η) is explained by the change in the quantum yield of the MoS$_2$ in the presence of plasmonic nanostructures. The PL quantum yield Y is defined by[53]

$$Y = \left(\frac{k_r}{k_r + k_{nr}}\right)$$

where, $k_r$ and $k_{nr}$ is the radiative and the nonradiative decay rates. The radiative decay rate is affected by the localized surface plasmonic fields whereas the nonradiative decay rate depends on plasmonic losses and exciton quenching.

The significant PL enhancement in both Ag - glass based nanoheterojunctions indicates an effective coupling between MoS$_2$ excitons and LSP resonances in nanostructures with large increase in the radiative decay rate. The plasmonic absorption transition dipole moment which is the collective oscillations of the surface electrons in AgPO$_3$ nanostructures that interacts with the transition dipole moments of MoS$_2$ (excitonic states of A and B) leading to collective states (Fig. 4b). Such states are often called hybrid states and result in stronger optical PL than the isolated states of the TMD. Since the plasmonic absorption band of 2.0eV in AgPO$_3$ (there are two bands, one at 2.0eV and the other at 2.5 eV) is in the vicinity of B- exciton transition of MoS$_2$ (2 eV) there is a higher probability for B-exciton plasmon dipole-dipole interaction due to the local field enhancement. The physical mechanism behind this process is illustrated in Fig. 4b. Altogether, the appearance in discrete and enhanced excitonic emission is led by the exciton (MoS$_2$) and surface plasmon (glass) coupling in the nanoheterojunctions (Glass:MoS$_2$) system. Further work is currently in progress to optimize this coupling via tuning of the Ag NPs size and fraction[26] into the AgPO$_3$ matrix.

**Discussion**

We have fabricated and demonstrated novel hybrid nanoscale heterojunctions of layered MoS$_2$ and metaphosphate glasses. The MoS$_2$ phase integrity and excitonic bands are preserved inside the glasses. The developed AgPO$_3$:MoS$_2$ composite heterojunctions exhibit a remarkably enhanced PL intensity with the presence of well-defined excitonic transitions. A strong modification of A- and B- exciton peak intensity by plasmonic nanostructure has been adopted. We have obtained a 6-fold enhancement factor for the intrinsically weak B exciton peak. Such enhancement factor for the B excitonic emission is explained with the help of dipole-dipole interaction via exciton-plasmon coupling. The ultrafast electron transfer process and carrier-carrier interaction in the nanoheterojunction system support the enhancement in the B excitonic emission. No doubt, the efficient dipole coupling of exciton-plasmon and tunability of B-excitonic emission find application in emerging valleytronic devices working with B excitons. Moreover, the presented fabrication process might be promising for large scale production of inexpensive nanophotonic, valleytronics and optoelectronic devices with tunable B excitonic emissions.

## Methods

### Sample preparation

*MoS$_2$ Nanoflakes*: MoS$_2$ flakes were prepared from bulk MoS$_2$ powder (grain size < 2 µm, Sigma Aldrich) using liquid phase exfoliation (LPE) method, as reported elsewhere.[54, 55] In detail, 40 mg of bulk MoS$_2$ powder was dissolved in 10 ml IPA. The solution was ultrasonicated for 60 min in a Elma S 30 H bath sonicator (Elma Schmidbauer GmbH, Germany) under 80 W power and 37 kHz frequency. Room temperature (<30ºC) was maintained throughout the exfoliation process (bath sonicator). After ultrasonication the dispersion was centrifuged to exclude the unexfoliated bulk MoS$_2$. The supernatant of the resulting dispersion was collected and used for subsequent experiments.

*AgPO$_3$ glass and AgPO$_3$:MoS$_2$ nanoscale heterojunction formation*: The development of AgPO$_3$:MoS$_2$ heterojunction glasses relies on the incorporation of MoS$_2$ flakes within silver metaphosphate glass (AgPO$_3$). First, a previously described procedure was followed for the preparation of the AgPO$_3$ glass substrates.[37, 38] Namely, equimolar amounts of high-purity AgNO$_3$ (99.995%) and NH$_4$H$_2$PO$_4$ (99.999%) dry-powders were melted in a platinum crucible. All weighing and mixing manipulations of the two powders were performed within a glove bag purged with dry nitrogen gas. After thorough mixing of the two powders, the melting batch was transferred to an electrical furnace initially held at 170 °C, while slowly heated up to 290 °C for the smooth removal of the volatile gas products. The furnace temperature was then raised to 450 °C and kept steady for 30 minutes, while performing frequent stirring in order to ensure melt homogeneity. AgPO$_3$ glasses were obtained in the form of 1 mm thick disk specimens with a diameter of around 10 mm, upon splat-quenching the melt. This well-established procedure results in AgPO$_3$ glasses with negligible water traces of less than 0.3 mol%, i.e. incapable of causing any optical or structural property modifications. Moreover, the glasses remain unaffected of room humidity (25-30%) for several months.

For the incorporation of MoS$_2$, the AgPO$_3$ glass substrate was positioned on a silicon wafer while a heating plate was employed in order to maintain a temperature around 80 °C. Ten drops of a previously prepared MoS$_2$ solution (0.76 mg/ml) were drop-casted on the surface of the AgPO$_3$ glass, while allowing 10 sec intervals between each drop in order to ensure smooth solvent vaporization. After solvent removal the residual MoS$_2$ flakes were randomly distributed on the AgPO$_3$ surface. Then, the temperature was raised to 170 °C for 2 minutes, i.e. 22 °C below the glass transition temperature of the AgPO$_3$ glass. At this temperature, the AgPO$_3$ glass becomes viscous and allows readily the smooth incorporation of the MoS$_2$ flakes within the glass matrix. Following MoS$_2$ immersion, the AgPO$_3$:MoS$_2$ nano-hybrid glass was splat-quenched between two silicon wafers, while instantly removed from the heating plate and left to cool down to room temperature. The employment of silicon wafers allows the formation of smooth surfaces on both sides of the composite glass specimens and renders them suitable for optical characterization. The $MoS_2$ incorporation process is presented in **Schematic S1**.

**Table 1. Description of the samples used in this study**

| Sample | Description |
| --- | --- |
| Bare $MoS_2$ | Isolated from bulk powder |
| $AgPO_3$ | Silver metaphosphate glass |
| Si/$MoS_2$ | $MoS_2$ on Si substrate |
| $AgPO_3$/$MoS_2$ | $MoS_2$ on the top of $AgPO_3$ |
| $AgPO_3$:$MoS_2$ | $MoS_2$ embedded into $AgPO_3$ |
| $NaPO_3$ | Sodium metaphosphate glass |
| $NaPO_3$:$MoS_2$ | $MoS_2$ embedded into $NaPO_3$ |
| $0.3AgI-0.7AgPO_3$:$MoS_2$ | $MoS_2$ embedded into $0.3AgI-0.7AgPO_3$ |

*Optical measurements*

The optical UV-Vis absorption spectra of the dispersion and solid films were carried out with a PerkinElmer, Lamda 950 UV/VIS/NIR spectrometer, USA. The Raman spectra were recorded under 473 nm laser excitation (Thermo Scientific) in the back-scattering geometry at ambient conditions at 300K. The Si substrate peak at 520 $cm^{-1}$ was used for calibration purposes.

For optical spectroscopy measurements, we used a Micro-Photoluminescence (μ-PL) setup and the spectra were collected in a backscattering geometry at 300K. As excitation source was used a continuous wave (CW) He-Ne 543 nm (2.28 eV) laser. An iHR-320 spectrometer (Horiba Scientific/Jobin Yvon Technology) equipped with a Syncerity multichannel charge-coupled device (CCD) Camera was employed to collect the spectra.

For the XRD measurements an X-Ray Rigaku (D/max-2000) diffractometer was employed, while being operated with a continuous scan of Cu Ka1 radiation with λ equal to 1.54056 Å. The morphology of the Ag NPs was studied by transmission electron microscopy (TEM, LaB6 JEOL 2100), after depositing drops of glass-powder/toluene solution onto a carbon-coated TEM grid. Finally, a field emission scanning electron microscope (SEM, JEOL, JSM-7000F) was used for the examination of the lateral dimension of dispersed 2D $MoS_2$ flakes, while atomic force microscopy (AFM) was employed to obtain the flakes' thickness (Digital Instruments with controller Nanoscope IIIa).

**Acknowledgements**

This work is supported by the project MoulTex, funded by EC framework programme HORIZON 2020, GA No-768705. The authors are grateful to A. Manousaki (IESL, FORTH) for her technical assistance with SEM studies, Lampros N. Papoutsakis (IESL, FORTH) for supervising XRD experiments, and A. Kostopoulou for capturing the HRTEM image. The Electron Microscopy Laboratory of the University of Crete is highly acknowledged for providing the HRTEM facility.

**Additional Information**

**Supplementary information** accompanies this paper at journal website.

**Competing financial interests:** The authors declare no competing financial interests.